\documentclass[12pt]{article}
\usepackage{graphics}
{\par\begingroup\addtolength{\leftskip}{#1}}%
{\par\endgroup}
\begin{document}
%
\noindent

\baselineskip 07mm

\def\lsun{{$L_{\odot}$} }
\def\amm{NH$_{3}$}
\def\teff{$T_{\rm{eff}}$}
\def\vmic{$v_{\rm{mic}}$}
\def\teff{$T_{\rm{eff}}$\ }
\def\vmic{$v_{\rm{mic}} $\ }
\def\afe{log $A_{\rm{Fe}} $\ }
\def\aa{${\it Astron.~Astrophys.}$\ }
\def\aal{${\it Astron.~Astrophys.~Lett.}$\ }
\def\apj{{\it Astrophys.~J.}\ }
\def\apjl{{\it Astrophys.~J.~Lett.}\ }
\def\aj{{\it Astron.~J.}\ }
\def\jqsrt{\it J.~Quant.~Spectrosc.~Rad.~Transf.\ }

\def\lteq{\mathrel{\hbox{\rlap{\hbox{\lower4pt\hbox{$\sim$}}}\hbox{$<$}}}}
\def\gteq{\mathrel{\hbox{\rlap{\hbox{\lower4pt\hbox{$\sim$}}}\hbox{$>$}}}}

\begin{center}

{\Large {\bf Water in K and M giant stars unveiled by ISO  
\footnote{Based on the data archives of ISO, an ESA project with 
instruments funded by ESA member states (especially the PI countries: 
France, Germany, the Netherlands, and the United Kingdom) 
and with the participation of ISAS and NASA.} 
}}

\vspace{8mm}

{\large  Takashi Tsuji}

\vspace{1mm}

Institute of Astronomy, The University of Tokyo, Mitaka, Tokyo, 181-0015 Japan

e-mail: ttsuji@ioa.s.u-tokyo.ac.jp


\vspace{25mm}

{\bf {\large abstract}}
\end{center}

{\large
Based on the spectra obtained with  Infrared Space Observatory, ISO,
we detected the 6.3\,$\mu$m bands of water in the late K giant Aldebaran 
($\alpha$ Tau) and several early M giant stars (between M0 and M3.5), 
which have been deemed to be too warm for tri-atomic H$_{2}$O molecule
to reside in their photospheres. The water column densities range 
$(0.2 - 2) \times10^{18}$\,molecules cm$^{-2}$ in our sample of K and M 
giant stars and the excitation temperatures are 1500\,K or higher. Thus, 
the water bands are  not originating in cool stellar winds either.  
The presence of water in the K and early M giant stars was quite unexpected
from the traditional picture of the atmosphere of the red giant star 
consisting of the photosphere, hot chromosphere, and cool wind.  We 
confirm that a rather warm  molecule forming region should exist as a new 
component of the atmosphere of  red giant stars and that this should be
a general phenomenon in late-type stars.}
 
 {\bf keywords}~~~ {Infrared: stars  -- molecular processes -- stars: 
atmospheres -- stars: chromospheres -- stars: individual: $\alpha$ Tau, $\beta$ And, 
$\alpha$ Cet, $\beta$ Peg, $\gamma$ Cru -- stars: late-type }
%

%
\vspace{5mm}
\clearpage

\section{Introduction}
A pioneering  attempt to observe stellar spectra from outside the 
Earth's atmosphere was undertaken more than 35 years ago with 
the balloon-borne telescope named Stratoscope II, and water was clearly 
detected in  Mira variables $o$ Cet and R Leo (Woolf et al. 1964).
Also, a possible presence of water in the normal M giants $\mu$ Gem 
(M3III) and $\rho$ Per (M4II) as well as in the early M supergiant  
$\alpha$ Ori (M2Iab) was suggested.  The presence of water in such 
non-Miras, however, was so unexpected at that time (and even
today) that it has not been understood correctly for a long time.
Instead, the absorption bands at  1.4 and 1.9\,$\mu$m attributed to
H$_{2}$O by the Stratoscope II observers were re-interpreted  as due to 
the CN red system which also has the bandheads at  1.4 and 1.9\,$\mu$m 
(Wing \& Spinrad 1970). In fact, it was known at that time that water 
can be observed only in the coolest M giant stars later than about M6 
(e.g. Johnson \& M\'endez 1970), while CN can be well observed in the 
warmer red giant stars. 
 
Further support for the proposition by Wing \& Spinrad (1970) was provided 
by the model photospheres of red giant stars developed at that time,
which  showed that water can  be abundant in M giants  with effective 
temperatures ($T_{\rm eff}$) lower than about 3200\,K and that 
$T_{\rm eff}$'s of M giants should be revised upward against the ones 
known at that time (Tsuji 1978). This upward revision of the effective 
temperature scale  was well consistent with the empirical scale based 
on the angular diameter measurements which showed $T_{\rm eff} \approx$ 
3250\,K for M6III (Ridgway et al. 1980). Thus, theory and observations 
that water can be observed only in M giants later than about M6 appeared 
to be consistent. This result further made it difficult to accept 
the Stratoscope II result that water could be detected in the M giant stars 
as early as M3-4.

Nevertheless, we recently found the possible presence of  water in the 
early M giant  $\beta$ Peg (M2.5II-III) (Tsuji et al. 1997) on the spectra we 
observed with the Short Wavelength Spectrometer, SWS (de Graauw et al. 1996), 
on board the ISO (Kessler et al. 1996).
This result was based on the analysis of the 2.7\,$\mu$m region where 
the H$_2$O $\nu_1$ and $\nu_3$ bands can be found, but the overlapping OH 
and CO bands made it difficult to clearly demonstrate the 
presence of the H$_2$O bands, especially by the low resolution spectrum
we had at that time. Also, possible presence of water in early M type stars was
suggested by the low resolution data obtained with IRTS (Infrared
Telescope in Space) of ISAS (Matsuura et al. 1999).
By these results, however, it might still be difficult to convince the 
presence of water in non-Mira stars earlier than about M6 against the 
general belief that water should not exist in such stars.
Now, it is possible to utilize a larger sample of  high resolution
ISO spectra recently released by ESA and, 
with the higher resolution, we detected the H$_2$O  $\nu_2$ bands 
in the 6.3\,$\mu$m region, where is little
disturbed by other  molecular bands. This observation finally
provides convincing evidence for the presence of water in normal red giants
including the late K and early M  giant stars.

\section{Detection of Water on ISO Spectra}

We used the spectra listed in Table 1 observed with the ISO SWS by its 
highest resolution grating mode, which gives a resolution of 
$R = \lambda/{\Delta \lambda} \approx 1600$ (FWHM$ \approx 188$\,km\,sec$^{-1}$).
The sample shown in Table 1 is probably all the red giants earlier than M4III
observed with the ISO SWS by this high resolution grating mode, even though more 
spectra were observed by the lower resolutions. Also, some spectra of late 
M giants were observed by the high resolution (e.g. Tsuji et al. 1997),
but we concentrate in this Letter to the case of red giant stars earlier 
than about M4 for which the presence of water is not clear yet.  
The spectra are reduced with the use of OSIA\footnote{
OSIA (Observers SWS Interactive Analysis) is a joint
development of the SWS consortium. Contributing institutes are
SRON, MPE, KUL and the ESA Astrophysical Division. } 
and the resulting spectra are shown in Fig.\,1. For comparison,
we  show in Fig.\,2  the spectra of H$_2$O in the form of 
log $B_{\nu}(T_{*}) e^{-\tau_{\nu}}$ with $\tau_{\nu} = 
\kappa_{\nu}(T) N_{\rm col}$, where $\kappa_{\nu}(T)$ is   
the absorption cross-section of H$_{2}$O  and  $ N_{\rm col}$ is 
the column density of H$_{2}$O (assumed to be $10^{18}$\,cm$^{-2}$ 
throughout). The spectra for $ T = 1000, 1500$, and 2000\,K are shown and 
some features to be used as signatures of  H$_2$O absorption are indicated 
by $ a - e $ in Fig.\,2.

The spectrum of  $\alpha$ CMa (A1Vm; $T_{\rm eff} \approx 10000$\,K) shown at 
the top of Fig.\,1 should show no stellar feature in this spectral region, and
the features shown may simply be noise  whose variations are
within about 0.01\,dex. ($\pm$1.2\%). The next two spectra of K giant stars 
$\alpha$ Boo (K1IIIb) and $\gamma$ Dra (K5III) may show some features which, 
however, do not agree with the signatures of water $a - e$ noted in Fig.\,2.
The features may be due to stellar CO lines (see Fig.\,3) and/or to noise. 
The spectrum no.\,4 of the K5 giant Aldebaran is quite different and
shows, if very weak, most signatures $a - e$ of water  againt noise. Also, 
the overall pattern of the spectrum of $\alpha$ Tau  is clearly different from 
that of the  spectra nos.\,1 - 3, which are rather similar to each other.
Then, the spectrum no.\,5 of the M0 giant $\beta$ And shows the H$_{2}$O 
signatures $a - e$  more clearly. The presence of water absorption in 
the spectra nos.\,6, 7, and 8 of $\alpha$ Cet (M1.5IIIa), $\beta$ Peg 
(M2.5II-III), and $\gamma$ Cru (M3.5III), respectively, is definite and 
we thus find convincing evidence for water in the early M giant stars. 
The water features are the strongest in $\beta$ Peg rather 
than in $\gamma$ Cru, the latest M giant in the present sample.
The identification of molecular absorption on stellar spectra is a
simple problem of pattern recognition, and the  presence of water
in $\alpha$ Tau can be convinced if we compare the spectrum no.\,4 with
the spectra nos.\,5 - 8.

\begin{table} 
\caption{{Program stars observed with the ISO SWS}}
            \begin{center}
            \begin{tabular}{l l r l l l}
            \hline
            \noalign{\smallskip}
    no. & object &  BS~ &  Sp. type  &  ${T_{\rm eff}}$(K) & ISO Obsno.\\ 
            \noalign{\smallskip}
            \hline 
            \noalign{\smallskip}
    1 & $\alpha$ CMa & 2491 & A1Vm       & $\approx 10000$     & 689 01202   \\
    2 & $\alpha$ Boo & 5340 & K1IIIb     & 4362$^{a}$          & 452 00101   \\
    3 & $\gamma$ Dra & 6705 & K5III      & 4095$\pm$163$^{b}$  & 040 02405   \\
    4 & $\alpha$ Tau & 1457 & K5III      & 3898$^{a}$          & 636 02102   \\
    5 & $\beta$ And  &  337 & M0IIIa     & 4002$\pm$178$^{b}$  & 795 01002   \\
    6 & $\alpha$ Cet &  911 & M1.5IIIa   & 3869$\pm$161$^{b}$  & 806 00924   \\
    7 & $\beta$ Peg  & 8775 & M2.5II-III & 3890$\pm$174$^{b}$  & 551 00705   \\
    8 & $\gamma$ Cru & 4763 & M3.5III    & 3626$^{a}$          & 609 00804   \\
            \noalign{\smallskip}
            \hline
            \end{tabular}
            \end{center}
\hspace{40mm}$^{a}$Cohen et al.(1996) ~~~$^{b}$Dyck et al.(1998)
\vspace{15mm}
\end{table}

The water spectra shown in Fig.\,2 are well sensitive to temperature, since
the features $a$ and $c$ are mainly contributed by the low excitation lines 
(typically L.E.P.$ < 2000$\,cm$^{-1}$) while the features $b, d,$ and $e$ by 
the higher excitation lines (L.E.P.$ > 2000$\,cm$^{-1}$). For this reason, 
the relative intensities of $b + d +e$ against $a +c$ are larger at higher 
temperatures. We notice that the observed spectra in Fig.\,1 do not agree 
with the trend of the predicted spectrum based on $T = 1000$\,K in Fig.\,2, 
and the excitation temperature of the water gas  in the observed red 
giants cannot be as low as 1000\,K. Instead, the relative intensities of 
the observed features appear to be more consistent with $T \approx 1500$\,K 
or somewhat higher. For evaluating water spectra shown in Fig.\,2, we used 
a calculated water linelist HITEMP (Rothman 1997), but its accuracy is 
unknown. Then, we also used a more extensive linelist by Schwenke \& 
Partridge (1997), and confirmed that the resulting spectra show little 
difference with those based on HITEMP at the resolution of Fig.\,2. This 
consistency of the available linelists is encouraging, although 
the accuracy of the linelists of hot water should be verified by 
laboratory data in future. Once temperatures can be known, the column 
densities can be estimated by comparisons of the observed and calculated 
water spectra. We found $N_{\rm col}$ between $2 \times 10^{17}$ 
($\alpha$ Tau) and $2 \times 10^{18}$ ($\beta$ Peg) cm$^{-2}$.

\section{Discussion }
In Fig.\,3, we examine  the predicted synthetic spectra based on the model 
photospheres, which are  essentially the same with those discussed before 
(Tsuji 1978) except that the photosphere is now assumed to be spherically 
symmetric rather than plane-parallel. Also, the opacity data are
largely updated by the use of the HITEMP linelist, whose log\,$gf$ values 
agree with those by Schwenke \& Partridge (1997) within 0.05\,dex.
The weak absorption features in the models of $T_{\rm eff}$ between 
4000 and 3400\,K are due to high excitation tails of the CO fundamental 
bands whose band origin is at 4.6\,$\mu$m. H$_{2}$O features appear 
first in  the model of  $T_{\rm eff}$ = 3300\,K and strengthen in 
the model of  $T_{\rm eff}$ = 3200\,K. On the other hand,
$T_{\rm eff}$'s of the early M giant stars (M0 - M3.5) are
between 3600 and 4000\,K as shown in Table 1. However, H$_{2}$O features
can never be predicted from the models of $T_{\rm eff} = 3600 - 4000$\,K.
Does this imply that our classical model photospheres are so useless? 

It is true that model photospheres of cool stars
are not yet perfect. However, stellar photospheres 
can be relatively well modeled based on few ad-hoc assumptions 
except possibly for the treatments of convection and turbulence,
and there is no reason why cool stars are exception only if molecular
opacities are properly taken into account. In fact, the present model 
photospheres of red giant stars have been tested by the
fact that the empirical effective temperature scale and the predicted one
based on our models show reasonable agreement as noted in Sect.\,1.
We believe that the photosphere of red giant stars can be modeled at
least approximately within the framework of the so-called classical
assumptions and that the model photospheres of cool stars
cannot be so wrong as to not able to predict the major molecular
features originating in the photosphere. However, we should notice that 
the stellar atmosphere, which represents all the observable outer layers, 
could not necessarily be represented by the model photosphere.  In other 
words, it should still be possible that some new component
remains unrecognized in the atmosphere of red giant stars beside
the known ones including the photosphere, chromosphere and wind. 

One possibility may be to assume the presence of  large starspots,
but such large starspots should give noticeable effects on other 
observables such as the spectral energy distributions, spectra, 
variabilities, activities etc. However, we know little evidence for 
such effects in the normal red giant stars.
Another possibility is to assume that the red giant stars are veiled
by a cloud of water vapor. In fact, we found clear evidence for such a case
in the M supergiant star $\mu$ Cep (M2Ia) by detecting the 
H$_{2}$O 6.3\,$\mu$m bands in emission on the ISO spectrum (Tsuji 2000b) 
and by confirming the 1.4 and 1.9\,$\mu$m bands in absorption on
the Stratoscope data (Tsuji 2000a). In another M 
supergiant star $\alpha$ Ori, the  H$_{2}$O 6.3\,$\mu$m bands appear in
absorption (Tsuji 2000b) and also absorption lines due to the
H$_{2}$O pure-rotation transitions were detected
by the high resolution ground-based spectroscopy (Jennings \& Sada 1998).  
The nature of water in the
red giant stars is rather similar to that in the red supergiant stars
(e.g. $T_{\rm ex} \approx 1500$\,K in the both cases), and we propose that the
similar model of a rather warm molecular sphere (MOLsphere) as for 
supergiants should be applied to the normal red giant stars.    

In this connection, it is interesting that the molecular cloud referred
to as ``CO-mosphere'' was found recently in the Sun by detecting CO 
emission beyond the solar limb (Solanski et al. 1994). Thus, the presence 
of the rather warm molecular sphere (MOLsphere) may be a common phenomenon 
in  late-type stars including the Sun, red giants and supergiants, and 
we hope that future detailed studies of the MOLsphere as well as of the 
CO-mosphere will clarify the physical basis of such a phenomenon.
Also, high excited water gas around very cool (super)giants has been
known from water masers for a long time (Knowles et al. 1969). 
But it now turns out that such warm water gas already exists in the late 
K and early M giants even though H$_{2}$O masers are not observed. 
This fact implies that the cradle for maser activity may have already
been germinating in K and M giant stars.

So far, the presence of the hot chromosphere ($T \approx 10^{4}$\,K) is 
known in K and M giant stars but no evidence for the solar-type corona 
(Linsky \& Haisch 1979). On the other hand, steady stellar wind already 
starts in K giant stars (Reimers 1977), but the origin of the wind is 
unknown yet. Recently, high sensitive infrared survey with the ISO (ISOGAL) 
revealed that efficient dust formation already starts in red giant stars 
with weak mass-loss rates (Omont et al. 1999). An interesting possibility 
is that the outer part of the MOLsphere is  cool enough for dust to form,
and this may explain why dust formation starts in the red giant stage 
prior to the AGB phase. Further, dust formed this way may be pushed 
outward by the radiation pressure and thus may explain the onset of 
the wind. This is of course not a solution to the origin of the dust 
and/or of the wind so long as the origin of the MOLsphere is unknown. 
But now it appears that the atmosphere of red giant stars is composed 
of the newly recognized MOLsphere in addition to the previously known 
photosphere, chromosphere and  wind. With this new component, 
a more unified picture and self-consistent theory for the atmospheric 
structure of red giant stars could be developed.

\section{Concluding Remarks}

The presence of water in the early M giants was once noticed more than 
35 years ago (Woolf et al. 1964), but this important discovery has 
not been understood properly and overlooked for a long time.
We had to wait ISO to confirm the presence of water in normal red giant 
stars  including the late K and early M giant stars.
By hindsight, this may partly be due to confusion not to have realized the 
difference of the photosphere and the atmosphere. 
For example, stellar photosphere could be modeled rather easily 
but it has often been referred to as a {\it model atmosphere}. 
However, what we had this way  is only a {\it model photosphere}
and we have no self-consistent {\it model atmosphere} 
of red giant stars yet. What Stratoscope II  suggested and what ISO  
finally unveiled is that the infrared 
spectra of  red giant stars involve a new problem that cannot be represented
 by the photospheric model, and that the atmosphere and the photosphere 
should be clearly distinguished.  
Certainly a more unified understanding of the stellar atmosphere 
should be required to properly interpret the infrared spectra of red 
giant stars.

\vspace{10mm}

\setlength{\parindent}{0 pt}

{\bf {\Large Acknowledgements}}
\vspace{2mm}

I thank I. Yamamura for making available his fine tools in applying OSIA,
T. Tanab\'e for helpful advice on OSIA, and the anonymous referee for 
valuable comments. This work is supported by the Grant-in-Aid for Scientific 
Research of the Ministry of Education \& Science no. 11640227.

\clearpage

\begin{figure}
\begin{center}
\resizebox{12cm}{!}{\includegraphics{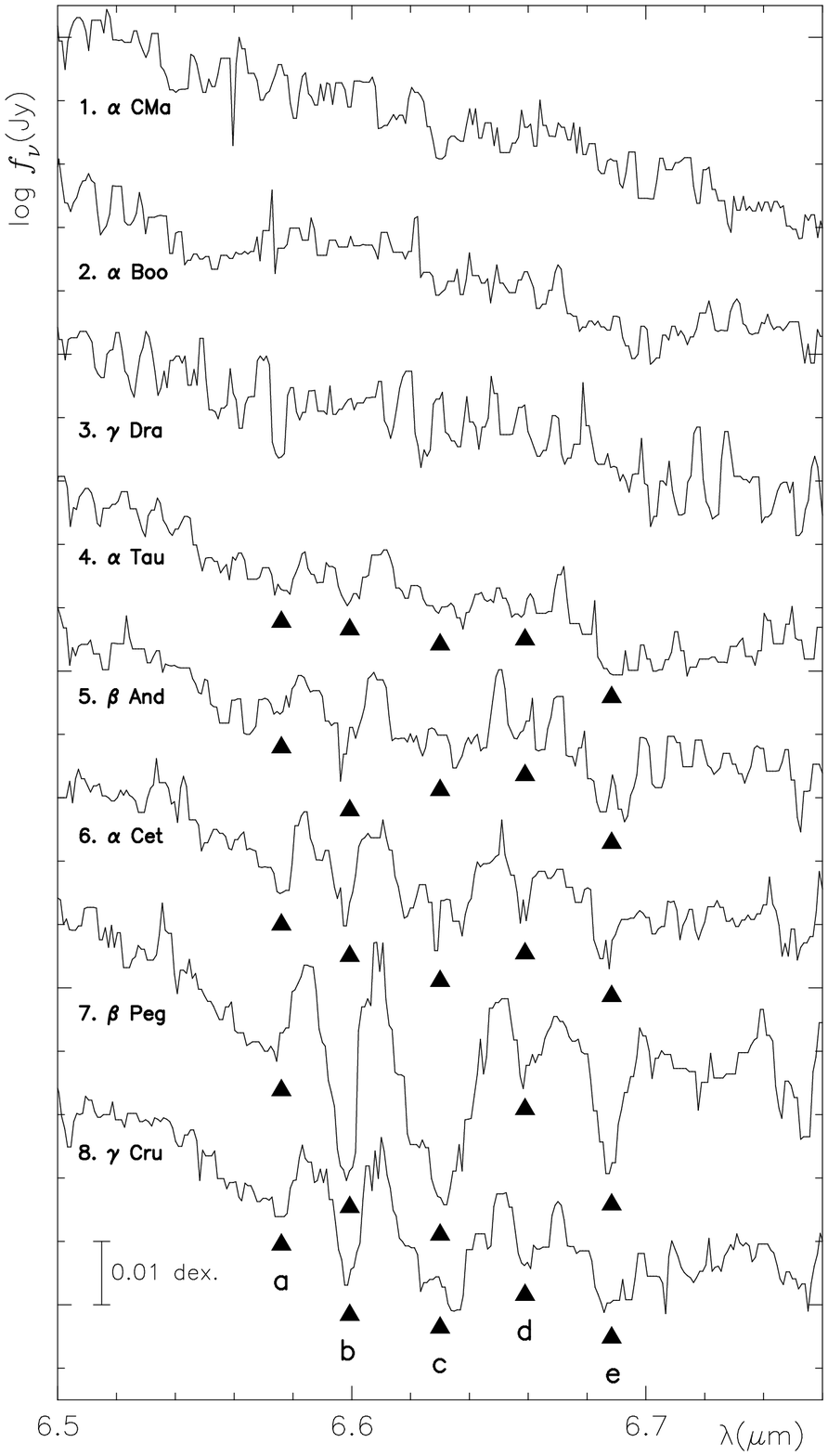}} 
\end{center}
\caption{
Spectra observed with the ISO SWS. The first three stars may serve as 
references in which no signature of water can be seen. The other five 
objects all show the signatures of water {\it a - e} predicted from 
the spectroscopic data of H$_{2}$O in Fig.\,2.
}
\end{figure}

\clearpage

\begin{figure}
\begin{center}
\resizebox{12cm}{!}{\includegraphics{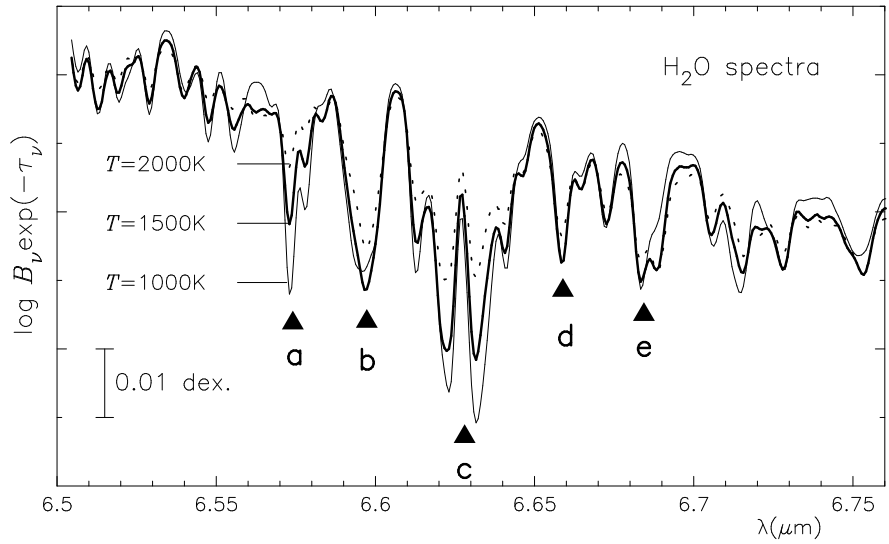}} 
\end{center}
\caption{
Spectra of water evaluated at high resolution ($R \approx 10^{5}$) and
convolved with the slit function of the ISO SWS (Gaussian with FWHM 
$\approx 188$\,km\,sec$^{-1}$) are shown for three  temperatures.  
The linelists by Rothman (1997) and by Schwenke \& Partridge (1997) 
are used and the results are almost the same. 
}
\end{figure}

\clearpage

\begin{figure}
\begin{center}
\resizebox{12cm}{!}{\includegraphics{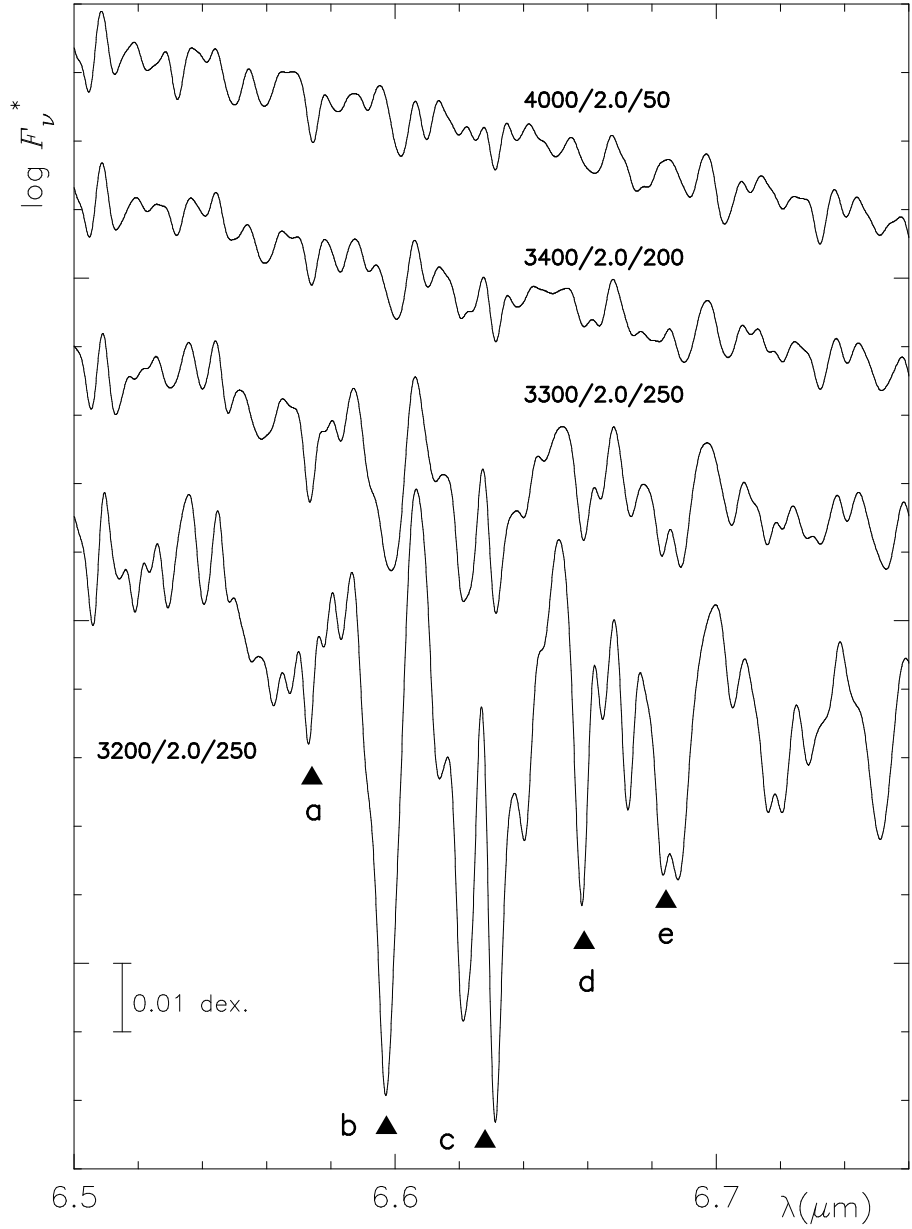}} 
\end{center}
\caption{
Predicted spectra by model photospheres whose basic parameters  
($T_{\rm eff}$/mass in unit of $M_{\odot}$/radius in unit of $R_{\odot}$)  
are indicated. The synthetic spectra are evaluated with the spectral 
line database including about a million of lines of CO, OH, CN, SiO, 
H$_{2}$O etc., and reduced to the  resolution of the observed spectra 
as in Fig.\,2.   
}
\end{figure}

\end{document}